\documentclass[prd,twocolumn,showpacs,preprintnumbers,amsmath,amssymb,floatfix,showkeys]{revtex4}
\usepackage{graphicx}% Include figure files
\usepackage{dcolumn}% Align table columns on decimal point
\usepackage{bm}% bold math
\def\bef{\begin{figure}}
\def\eef{\end{figure}}

\newcommand{\be}[1]{\begin{equation}\label{#1}}  
\newcommand{\beq}{\begin{equation}}
\newcommand{\eeq}{\end{equation}}
\newcommand{\bsub}{\begin{subequations}\begin{eqnarray}}
\newcommand{\esub}{\end{eqnarray}\end{subequations}}
\newcommand{\bea}[1]{\begin{eqnarray}\label{#1}}
\newcommand{\eea}{\end{eqnarray}}
\newcommand{\bd}{\begin{displaymath}}
\newcommand{\ed}{\end{displaymath}}

\newcommand{\slax}[1]{\not{\! {#1}}}
\newcommand{\eq}[1]{Eq.~(\ref{#1})}
\newcommand{\eqs}[1]{Eqs.~(\ref{#1})}
\newcommand{\eqss}[2]{Eqs.~(\ref{#1}), (\ref{#2})}
\newcommand{\fig}[1]{Fig.~(\ref{#1})}

\newcommand{\gev}{\,\mathrm{GeV}}
\newcommand{\evs}{\,\mathrm{eV}^2}

\def\ijmp{Int.\ J.\ Mod.\ Phys.\ }

\def\jhep{J.\ High Energy Phys.\ }

\def\np{Nucl.\ Phys.\ }		%% \textbf{Bxxx}
\def\pl{Phys.\ Lett.\ }  %% up to 169B (1986)
\def\plb{Phys.\ Lett.\ B\ }	%% from 171 (1986)
\def\prep{Phys.\ Rep.\ }
\def\prl{Phys.\ Rev.\ Lett.\ }

\def\sjnp{Sov.\ J.\ Nucl.\ Phys.\ }
\def\etal{\textit{et al.}}

\newcommand{\dgamma}{\Delta \gamma}
\newcommand{\barphi}{\bar{\phi}}
\newcommand{\ra}{\rightarrow}
\newcommand{\equi}{\mathrm{eq}}
\newcommand{\s}{\mathrm{S} }
\newcommand{\trc}{\mathrm{Tr}}
\newcommand{\nequi}{n_\equi}

\newcommand{\tstar}{T_\star}
\begin{document}

\preprint{CFNUL/04-LB1}
%\preprint{}

\title{Neutrino helicity asymmetries in leptogenesis}

\author{Lu{\'{\i}}s Bento}
\email{lbento@cii.fc.ul.pt}
\author{Francisco C.\ Santos}
\email{fsantos@cii.fc.ul.pt}
\affiliation{%
 Universidade de Lisboa, Faculdade de Ci\^ encias
 \\ Centro de F\'{\i}sica Nuclear da Universidade de Lisboa \\
Avenida Prof. Gama Pinto 2, 1649-003 Lisboa, Portugal
}

\date{April, 2005}

%%\maketitle

\begin{abstract}
It is pointed out that the heavy singlet neutrinos characteristic of leptogenesis develop asymmetries in the abundances of the two helicity states as a result of the same mechanism that generates asymmetries in the standard lepton sector.
Neutrinos and standard leptons interchange asymmetries in collisions with each other.
It is shown that an appropriate quantum number, $B-L'$, combining baryon, lepton and neutrino asymmetries, is not violated as fast as the standard $B-L$.
This suppresses the washout effects relevant for the derivation of the final baryon asymmetry.
One presents detailed calculations for the period of neutrino thermal production in the framework of the singlet seesaw mechanism.
\end{abstract}

% PACS, the Physics and Astronomy
                             % Classification Scheme.
%Use showkeys class option if keyword
                              %display desired

\maketitle

% Comment out if separate title page not required
%%\thispagestyle{empty}

\newpage
%%\setcounter{page}{1}

%%%%%%%%%																															%%%%%%%%%%%%%%%%%%%
%%%%%%%%%																															%%%%%%%%%%%%%%%%%%%
%%%%%%%%%																															%%%%%%%%%%%%%%%%%%%
%%%%%%%%%																															%%%%%%%%%%%%%%%%%%%
\section{Introduction}\label{introduction}

Leptogenesis is an attractive way of generating the baryon number of the 
Universe~\cite{Buchmuller00}. 
The main idea put forward by Fukugita and Yanagida~\cite{Fukugita86,Luty92,Covi96}
is that a lepton number asymmetry can be produced in the decays of heavy singlet neutrinos into leptons and Higgs bosons and such an asymmetry 
is partially transferred to the baryon sector through electroweak sphaleron processes~\cite{Kuzmin85}
that violate $B$ and $L$ but not $B-L$.
The mechanism requires nonconservation of lepton number and $CP$ provided by 
neutrino Majorana masses and complex Yukawa couplings.
Both masses and couplings form the well known singlet seesaw model \cite{Minkowski77,GellMann79,Yanagida79,Glashow79,Mohapatra80}
%%\cite{Minkowski77} - \cite{Mohapatra80}
of light neutrino masses and thus establish a close relationship between baryogenesis and 
low energy phenomenology 
\cite{Nelson90,Fischler91,Buchmuller93,Pilaftsis97,Hirsch01,Buchmuller01,%%
Endoh02,Branco02,davidson02,Ellis02,Buchmuller02,davidson03,Rebelo03,%%
Ibarra04,Giudice04,Hambye04}.
%%%%%%\cite{Nelson90} - \cite{Hambye04}.
This connection contributed for the present wide interest in leptogenesis.

The calculation of the final baryon asymmetry has been done in the literature with increasing levels of accuracy 
\cite{Pilaftsis97,Hirsch01,Buchmuller02,%%
Giudice04,Hambye04,Plumacher97,Flanz98,Barbieri00,Pilaftsis04}
%%\cite{Pilaftsis97,Hirsch01,Buchmuller02}, 
%%\cite{Giudice04} - \cite{Pilaftsis04}
 but the main elements have remained the following.
Singlet neutrino reactions are not symmetric under $CP$ due to nontrivial complex Yukawa couplings in the neutrino mass eigenstate basis.
Departure of neutrino densities from thermal equilibrium values are a necessary condition~\cite{Sakharov67}
to obtain net lepton asymmetry sources.
This occurs if neutrinos are not produced in the inflaton decay but only gradually from active lepton and Higgs boson collisions.
It occurs also to some extent when any of the neutrino species undergoes the transition to the respective nonrelativistic temperature epoch.
Weak sphaleron processes transform a fraction of the generated $B-L$ asymmetry into baryon number.
The final $B-L$ and baryon asymmetries depend on lepton number violating reactions whose net effect is to dissipate $B-L$.
They include the reactions 
$\barphi \, \barphi \ra l_i  l_j$, $\,\barphi \, \bar{l}_j \ra \phi \, l_i $, 
$\,\bar{l}_i \bar{l}_j \ra \phi \, \phi$,
that violate lepton number by two units, but also processes that violate standard lepton number by one unit~\cite{Luty92} 
such as top quark and electroweak gauge boson scatterings like
$\bar{t}\, q_t \ra N_a l_i$, $\, \barphi \,W \ra N_a l_i$ and $ \barphi \,B \ra N_a l_i$.
	Neutrino number densities and the set of standard lepton, quark and Higgs boson number asymmetries obey a system of coupled Boltzmann equations that is necessary to integrate to derive the final (present) $B-L$ asymmetry.

And here comes the main point of this paper. 
This system is incomplete because, contrary to what has ever been assumed, the two helicity states of singlet Majorana neutrinos do not have exactly the same abundances.
We make the case that leptogenesis mechanisms naturally generate asymmetries in the two neutrino helicity state abundances, in the same way as the standard lepton asymmetries, and that the neutrino helicity asymmetries play a role in the system of Boltzmann equations that govern the evolution and transport of particle asymmetries, lepton number in particular, and therefore contribute to the determination of the final $B-L$ asymmetry as a function of the fundamental parameters of the theory.
We present explicit calculations in the framework of the singlet seesaw mechanism.

In next section we review some of the properties of this model that are relevant for leptogenesis in relation with the light neutrino mass spectra~\cite{Bahcall04,Maltoni04,Goswami04} 
as indicated by solar 
\cite{Davis04,Cleveland98,Hampel99,Cattadori02,Abdurashitov02,Fukuda02,Ahmad02},
%%\cite{Davis04} - \cite{Ahmad02}, 
atmospheric \cite{Fukuda98,Ambrosio03,Sanchez03}
and terrestrial neutrino experiments~\cite{Eguchi03,Ahn03}.
In section \ref{helicity} we introduce the concept of neutrino helicity asymmetries and discuss their relevance for the transport of lepton number.
An appropriate quantum number $B-L'$ is proposed to replace the usual $B-L$ difference.
In section \ref{generation} we calculate explicitly the lepton and neutrino asymmetries generated during the phase of singlet neutrino thermal production.
This was presented in a brief fashion in a meeting~\cite{beyond03}.
Here we give a complete account of the work and improve the integrations over phase space by including Pauli blocking and lepton thermal mass effects in the numerical calculations.
The Higgs boson thermal mass had already been taken into account.
In section \ref{washout} we study the washout processes and evaluate the damping rate of $B-L'$.
We compare with the traditional treatment without neutrino helicity asymmetries and take the appropriate lessons.
The main results are summarized in the final section.

%%\newpage
%%%%%%%%%																															%%%%%%%%%%%%%%%%%%%
%%%%%%%%%																															%%%%%%%%%%%%%%%%%%%
%%%%%%%%%																															%%%%%%%%%%%%%%%%%%%
%%%%%%%%%																															%%%%%%%%%%%%%%%%%%%
\section{Seesaw model}\label{seesaw}

The singlet seesaw mechanism~\cite{Minkowski77}
adds to the standard model singlet (left-handed) neutrinos,
$N_a$,
with heavy Majorana masses and Yukawa couplings with the standard lepton and Higgs doublets,
$l_i$ and $\phi$, of the form
\be{Yuk} 
h_{ia}l_i N_a \phi + \frac{1}{2} M_{a} N_a N_a  
+  {\rm H.C.}  \; .
\eeq
%%(\ref{Yuk}) %%%%%%%%%%%%%%%%%%%%%%%%%%%%%%%%%%%%%%%%%%%%%%
Spontaneous breaking of SU(2) $\times$ U(1) yields the light neutrino mass matrix
($v= \langle \phi^0 \rangle$)
\be{mij}
m_{ij} =- (h M^{-1} h^{T} )_{ij} \, v^{2}
			\; .
\eeq
%%(\ref{mij}) %%%%%%%%%%%%%%%%%%%%%%%%%%%%%%%%%%%%%%%%%%%%%%
%where $v= \langle \phi^0 \rangle$.
The proper decay rate of $N_a$ into leptons and Higgs is given by
%% $N_a \rightarrow l\phi,\bar{l},\bar{\phi}$ 
\be{gna}
\Gamma_{N_a}^0 = \frac{(h ^{\dagger} h)_{aa}}{8\pi} M_a
			\; ,
\eeq
if one ignores thermal effects.
Delayed decay occurs when the ratio to the Hubble expansion rate $H$ at the temperature $T=M_a$, 
\be{ka}
K_a = \frac{\Gamma_{N_a}^0}{H_{(T=M_a)}} %%{H_{M_a}} 
			\; ,
\eeq
is small.
In the radiation era, 
$H = 1.66\, g_\ast^{1/2}T^2/M_{P}$, 
where $g_\ast$ denotes the number of relativistic degrees 
of freedom, 107.5 in the standard model.
It is enough to compare the sum
%%$K =\sum K_a$,
\be{k}
K = \sum K_a =
 (10^3 \; \mathrm{ eV}^{-1} ) \, \trc [  h M^{-1} h ^{\dagger} ] v^2 
\end{equation}
%%(\ref{k}) %%%%%%%%%%%%%%%%%%%%%%%%%%%%%%%%%%%%%%%%%%%%%%%%%%%%%%%%%%%%%%
with the light neutrino mass scale 
$\trc [ m ]$
to conclude that the delayed decay condition, $K_a <1$, is in general in conflict with the atmospheric neutrino mass gap~\cite{Fukuda98},
$\Delta m^2 \approx 2.5 \cdot 10^{-3} \evs$,
which implies $K > 50$. 
Strictly speaking, the delayed decay scenario only requires that the lightest of the heavy neutrinos satisfies $K_a <1$. 
But that is not the most natural picture, in particular if light neutrinos are 
quasidegenerate.
In section \ref{generation} we will assume that all parameters $K_a$ are large, of the order of 50 or more, which has the effect that singlet neutrinos enter in thermal equilibrium at relativistic temperatures $T_a \gg M_a$. 

%%\newpage
%%%%%%%%%																															%%%%%%%%%%%%%%%%%%%
%%%%%%%%%																															%%%%%%%%%%%%%%%%%%%
%%%%%%%%%																															%%%%%%%%%%%%%%%%%%%
%%%%%%%%%																															%%%%%%%%%%%%%%%%%%%
\section{Neutrino helicity asymmetries}\label{helicity}

Sterile neutrino Yukawa couplings $l_i N_a \phi$ conserve total lepton number, $L_T$, assigned as $L=-1$ for left-handed neutrino fields, $N_a$, and $L=+1$ for the right-handed neutrino conjugate fields $\bar{N}_a$.
Neutrino Majorana masses break lepton number but it is clear that in the ultrarelativistic limit the masses are negligible and  neutrinos and antineutrinos may have unequal abundances and symmetric chemical potentials like any other particles.
In that case neutrinos carry lepton number that can be exchanged between them and standard leptons in collisions mediated by Yukawa interactions.
The Majorana masses do not change this completely.
One faces a similar problem at defining lepton number of Majorana mass solar or atmospheric neutrinos, or lepton asymmetries of neutrinos at the time of Big Bang Nucleosynthesis~\cite{Langacker82}.
The main difference is that they deal with standard active neutrinos and not necessarily sterile neutrinos.
Free massive neutrino states are solutions of the Dirac equation that can be discriminated as spin eigenstates and in particular as helicity eigenstates. 
The $\pm 1/2$ helicity eigenstates spinors satisfy the relation
\be{ub}
u \, \bar{u} = \frac{1}{2}
\left(1 \pm \gamma_5 \frac{E-M\, \gamma^0}{p}\right) (\not{p}+M)
	\; ,
\eeq
where $E$ is the neutrino energy, $p$ the momentum and $M$ the mass.
When neutrinos are ultrarelativistic, helicity and chiral states are almost identical
and their lepton number is maximal ($\pm 1$).
For an arbitrary neutrino state the lepton number current density is evaluated as the expectation value
\be{jmul}
J^\mu_L = \left\langle \bar{N} \gamma^\mu \gamma_5 N \right\rangle
= \frac{1}{2} \left\langle \bar{\chi} \gamma^\mu \gamma_5 \chi \right\rangle
	\; ,
\eeq
where $N$ is either a left-handed chiral field $N_a$ or its conjugate $\bar{N}_a$, and 
$\chi = N_a + \bar{N}_a$ is a Majorana field (for the quantization of Majorana mass fields see refs.~\cite{Case57,Schechter80,Bento00}).
As a result a positive (negative) helicity eigenstate carries a well defined average lepton number equal to the neutrino speed $v=p/E$ ($-v$):
\be{L}
L = \frac{u^\dagger \gamma_5 u}{u^\dagger  u} = \pm v
	\; .
\eeq
The lepton number vanishes in the nonrelativistic limit due to the Majorana nature of the neutrino mass.

One may observe that in contrast to lepton number, helicity is not invariant under Lorentz transformations.
However, in an isotropic Universe the comoving thermal bath frame is a privileged  frame where isotropy enforces the spin density matrix to be diagonal in the helicity basis.
That means that each of the neutrino flavors $N_a$ can be divided in two populations of opposite helicities and well defined distribution functions $f_a^\pm$. 
The total lepton number carried by each neutrino species is equal to
\be{la}
L_a = V \int \frac{d^3p_a}{(2\pi)^3} (f_a^+ - f_a^-)\, v_a
	\, ,
\eeq
where $v_a$ is the neutrino speed and $V$ the spatial volume.
It will prove convenient to work instead with the helicity asymmetries i.e., the differences between positive and negative helicity neutrino abundances,
\be{lambdaa}
\Lambda_a = V \int \frac{d^3p_a}{(2\pi)^3} (f_a^+ - f_a^-)
	\, .
\eeq
Sterile neutrinos have been deprived of lepton number in the leptogenesis literature 
%%(see e.g.~\cite{Buchmuller00,Luty92} )
because they have Majorana masses (one exception is the oscillation mechanism of ref.~\cite{Akhmedov98}).
But contrary to the lepton number assignment, that is to an extent arbitrary, it is a unambiguous fact that neutrinos develop helicity asymmetries as a result of collisions with standard leptons and Higgs and also directly from the same leptogenesis processes that generate lepton asymmetries.
That is shown explicitly in section \ref{neutrino}.

It is important to realize that neutrino helicity asymmetries affect the way lepton number is transported in collision processes and consequently the so-called lepton number washout effect. 
Take for example 
%% the decays 
%% $N_a \leftrightarrow l_i \phi$, $N_a \leftrightarrow \bar{l}_i \bar{\phi}$ and 
scatterings like
$N_a l_i \leftrightarrow \bar{t}q_t$, $N_a \bar{l}_i \leftrightarrow t \bar{q}_t$,
and crossed channels, that violate standard lepton number by one unit
($t$ is the right-handed top quark and $q_t$ the quark iso-doublet linear combination that has a Yukawa coupling with it).
The correct evaluation of the lepton number violation rate must take into account that the reaction rates depend on the neutrino helicity and the two helicity states have different abundances.
Reactions with positive helicity neutrinos in initial or final states like $N_a^+ l_i$ are suppressed with respect to the opposite helicity states $N_a^- l_i$.
The former are possible only because neutrinos have Majorana masses and their transition amplitudes are suppressed by the Lorentz contraction factors $M_a/E$.
If neutrinos are in thermal equilibrium their helicity asymmetries can be parametrized with degeneracy parameters $\eta_a = \mu_a/T$ as any other fermions,
$\eta_a$ for positive helicity and $-\eta_a$ for negative helicity states, and
the rate of standard lepton number violation contains terms like
\bea{lrate}
\gamma_{N_a l_i \ra \bar{t}q_t} - \gamma_{\bar{t}q_t \ra N_a l_i} & =& 
	{\gamma}_+ (\eta_\phi + \eta_{l_i} + \eta_a)  \nonumber \\ &+ &		
 {\gamma}_- (\eta_\phi + \eta_{l_i} - \eta_a) 	
	\;, 
\eea
where $\gamma_+$, $\gamma_-$ stand for the average reaction rates of
$N_a^+ l_i \leftrightarrow \bar{t}q_t$ and
$N_a^- l_i \leftrightarrow \bar{t}q_t$ respectively. %%, and
	We have replaced $\eta_t$ and $\eta_{q_t}$ with the Higgs degeneracy parameter using
the constraint $\eta_t - \eta_{q_t} = \eta_\phi$ 
enforced by the rapid top quark Yukawa interactions.
	Identical expressions apply to other single neutrino absorption and emission reactions, decays and inverse decays $N_a \leftrightarrow l_i \phi$
(at temperatures $T \lesssim M_a$ where the Higgs and lepton thermal masses are small enough),
or $\barphi \leftrightarrow N_a l_i$
(at temperatures $T \gtrsim 2 M_a$), and respective radiative processes with one electroweak gauge boson in the initial or final state.
	Notice that because $\gamma_+$ is smaller than $\gamma_-$ the result depends on the neutrino chemical potentials.
The two rates coincide with each other only when neutrinos are at rest.
On the other hand one should not expect that the neutrino degeneracy parameters $\eta_a$ are damped faster by the lepton number violating reactions than the combinations 
$\eta_\phi + \eta_{l_i}$, also damped by $\Delta L =2$ collisions like 
$\phi l_i \leftrightarrow \bar{\phi} \bar{l}_j$.

Contrary to the partial lepton numbers of standard leptons, $L_i$, and neutrinos, $L_a$, the total lepton number
\be{l}
 L_T = \sum_i L_i + \sum_a L_a
	\; 
\eeq 
and the quantum number $B-L_T$ are  conserved by neutrino-lepton Yukawa couplings. 
In addition, $B-L_T$ is conserved by sphalerons and is only violated by neutrino Majorana masses.
For practical purposes namely, application of the constraints imposed by $CPT$ invariance and unitarity on the collision rates, it is more convenient to work with the helicity asymmetries $\Lambda_a$ rather than the lepton numbers $L_a$ because the states with definite distribution functions are helicity eigenstates, not (chiral) lepton number eigenstates.
%%, and $\Lambda_a$ are directly related with the helicity state abundances.
On the other hand, the quantum number
\be{lprime}
 L' = \sum_i L_i + \sum_a \Lambda_a
	\; 
\eeq 
has essentially the same interesting properties as the total lepton number $L_T$.
The combination $B-L'$, also conserved by weak sphalerons, reduces to the standard model $B-L$ when the heavy neutrinos vanish from the Universe.
It is conserved by neutrino Majorana masses but its violation by Yukawa couplings is suppressed by the neutrino masses i.e., mass over energy ratios.

Neutrino helicity asymmetries play a role in the transport of standard lepton number.
We have shown for instance that processes that violate flavor quantum numbers $L_i$ and $\Lambda_a$ by one unit contribute to the violation rate of $L_i - B/3$ as follows
(if all particles are in thermal equilibrium):
\bea{dlimbdt}
\frac{d(L_i - B/3)}{dt} &=&
 - \gamma_{ia}^{(0)} (\eta_\phi + \eta_{l_i} - \eta_a)  \nonumber \\
& & -{\gamma}_{ia}^{(2)} (\eta_\phi + \eta_{l_i} + \eta_a) 
 +  \cdots
%%	\; ,
\eea
where $\gamma_{ia}^{(0)}$ is the total rate of $L'$ conserving reactions and $\gamma_{ia}^{(2)}$ the total rate of $\Delta L' = \pm 2$ reactions.
	In turn, the standard lepton asymmetries are transferred to the neutrino sector in collisions with them.
	The same processes as above contribute to the violation rates of helicity asymmetries $\Lambda_a$ as
\be{dladt}
\frac{d \Lambda_a}{dt} =  \gamma_{ia}^{(0)} (\eta_\phi + \eta_{l_i} - \eta_a) 
- {\gamma}_{ia}^{(2)} (\eta_\phi + \eta_{l_i} + \eta_a) + \cdots
	\;.
\eeq
This proves that the neutrino helicity asymmetries cannot be assumed identically zero because the $L'$ conserving reactions are faster than the $L'$ violating ones.
It is also clear that only the later contribute to the violation rate of $B-L'$.
We will return to this point in section \ref{washout}.

This discussion shows that in order to correctly evaluate the lepton number washout effects neutrino helicity asymmetries are an essential ingredient.
They have been so far completely ignored in the leptogenesis literature.
The calculation of the lepton number generated during the decay phase of the lightest neutrino(s) and integration of the Boltzmann equations including neutrino helicity asymmetries is complicated by the fact that the neutrinos are neither purely nonrelativistic nor ultrarelativistic at temperatures close to their masses.
In this paper we limit ourselves to the temperature range where neutrinos are ultrarelativistic which permits first order calculations on their mass over temperature ratios.
In next section we calculate the $B-L'$ asymmetry generated during neutrino thermal production.

%%\newpage
%%%%%%%%%																															%%%%%%%%%%%%%%%%%%%
%%%%%%%%%																															%%%%%%%%%%%%%%%%%%%
%%%%%%%%%																															%%%%%%%%%%%%%%%%%%%
%%%%%%%%%																															%%%%%%%%%%%%%%%%%%%
\section{Generation of lepton asymmetries} \label{generation}

%%%%%%%%%																															%%%%%%%%%%%%%%%%%%%
%%%%%%%%%																															%%%%%%%%%%%%%%%%%%%
%%%%%%%%%																															%%%%%%%%%%%%%%%%%%%
%%%%%%%%%																															%%%%%%%%%%%%%%%%%%%
\subsection{Neutrino thermal production} \label{production}

Let us now examine the leptogenesis processes in detail during neutrino thermal production.
We assume that the Universe is initially empty of singlet neutrinos,
not produced in the inflaton decay
but only thermally from standard leptons and Higgs.
	The dominant thermalizing reactions are identical to the ones of the charged lepton isosinglets~\cite{Cline93}, namely, top quark scattering processes like
$q_t \bar{t}_R \rightarrow l_i N_a$,
Higgs boson decay 
$\barphi \rightarrow l_i N_a$
and related scattering processes with one additional electroweak gauge boson in the initial or final state.
	Higgs boson decays into leptons and neutrinos, first considered in ref.~\cite{beyond03}
in the context of leptogenesis, are allowed as much as the decays
into a lepton iso-doublet and a charged lepton isosinglet,
$\barphi \rightarrow l_i e_i$,
because the Higgs has a significant thermal mass, 
$m_\phi = x_\phi T \sim 0.6 \, T$,
larger than the lepton thermal masses~\cite{Klinov82,Weldon82a,Davidson94}
as pointed out in ref.~\cite{Cline93}.

Let $n_a$ denote the average neutrino number densities per helicity degree of freedom and 
$Y_a = n_a V$ the abundances in a fixed comoving volume whose  spatial volume $V$ expands as 
$ \dot{V} = 3 H\, V$.
For definiteness we assume that the reheating temperature is much higher than the neutrino masses so that neutrinos thermalize while they are ultrarelativistic.
Then, the equilibrium densities and abundances are equal to 
$\nequi = 0.90\, T^3/\pi^2$ 
and $Y_\equi = \nequi V$, respectively.
	Assuming for simplicity that the distribution functions scale with the equilibrium distribution functions $f_a^\equi$ as
\be{fa}
f_a =  \frac{n_a}{\nequi} f_a^\equi
	\;,
\eeq
and approximating the Pauli blocking factors $1 - f_a$ with $1-f_a^\equi$ in neutrino emission reactions, the neutrino abundances evolve as
\be{Ya}
\dot{Y}_a = \Gamma_{a} (Y_\equi - Y_a)
	\; .
\eeq
Here we neglect reactions with two or more neutrino states that are of higher order in the Yukawa couplings.

$\Gamma_{a}$ represent the neutrino collision frequencies.
At temperatures much higher than the neutrino masses $\Gamma_{a}$ scale with the temperature as follows:
\be{ga}
\Gamma_{a} = \frac{\beta}{8 \pi} (h^\dagger h)_{aa} T
	\; .
\eeq
They are comparable with the 
$\barphi \rightarrow l_i N_a$
proper decay rate, equal to
$\frac{1}{16\pi}$ $(h ^{\dagger} h)_{aa} m_\phi$
if one ignores thermal effects other than the Higgs boson mass.
The coefficient $\beta$ gets contributions from the Higgs decay, top quark and $W$, $B$ gauge boson scatterings in analogy with the charged leptons case~\cite{Cline93,Giudice04}.
The relative weights of these reactions depend on the temperature because the couplings constants~\cite{Ford93}
 and thermal mass factors~\cite{Klinov82,Weldon82a,Davidson94,Weldon82b} run with the energy scale~\cite{Giudice04}.
Using $m_\phi = 0.6\, T$ for the Higgs mass and $10^{9}$ GeV temperature scale coupling constants namely, 
$\alpha_s \approx 1/26$, $\alpha_{\mathrm{w}} \approx 1/38$, $\alpha' \approx 1/81$
for the strong, SU(2) and U(1)$_Y$ gauge interactions respectively,
and $\lambda_t \approx 0.60$ for the top Yukawa coupling,
one obtains $\beta \approx 1/7$, 
where $46 \%$ comes from the Higgs boson decay, 
$41 \%$ from $W$ and $B$ electroweak gauge boson scatterings, 
and $13 \%$ from right-handed top quark scatterings.

In the radiation era the Hubble expansion rate scales as $H= 1/2t \propto T^2$ and 
the assumed initially zero neutrino densities $n_a$ converge exponentially to the equilibrium densities:
\be{na}
n_a = \nequi \left( 1 - e^{-T_a /T} \right)		\; .
%T^{-1}} \right)		\; .
\eeq
%%(\ref{na}) %%%%%%%%%%%%%%%%%%%%%%%%%%%%%%%%%%%%%%%%%%%%%%%%%%%%%%%%%%%%%%
	The relaxation temperatures $T_a$ are given as
\be{Ta}
T_a = \frac{\Gamma_{a} T}{H} = \beta \, K_a M_a
	\; ,
\eeq
%%(\ref{Ta}) %%%%%%%%%%%%%%%%%%%%%%%%%%%%%%%%%%%%%%%%%%%%%%%%%%%%%%%%%%%%%%
and the second identity establishes a relation with the parameters $K_a$ of \eq{gna} that control the speed of neutrino decays when they become nonrelativistic and vanish from the Universe.
The low energy neutrino data indicates that the parameters $K_a \gtrsim 50$ are large, see \eq{k}.
As a result singlet neutrinos reach thermal equilibrium at temperatures $T_a \gtrsim 10 M_a$ when they are still ultrarelativistic.

We have defined neutrino densities in the basis of Majorana mass eigenstates as usual.
	However, this is not valid for all temperature scales because neutrinos also get thermal masses from the interactions with the lepton - Higgs thermal bath.
	The chiral conserving thermal mass terms are~\cite{Klinov82,Weldon82a}
\be{m2ab}
m^2_{ab} = \frac{1}{8} (h^\dagger h)_{ab} T^2
	\; .
\eeq
At high enough temperatures the thermal masses dominate over the vacuum masses $M_a$ and the neutrino Hamiltonian eigenstates are eigenstates of the matrix $(h^\dagger h)_{ab}$.
The opposite happens when $M_a$ are much larger than the thermal masses, and 
at temperatures where the two types of masses are comparable with each other the neutrinos undergo strong flavor oscillation processes.
One can estimate the thermal over vacuum mass ratios using \eqss{ka}{Ta}:
\be{mm}
\frac{m^2_{aa}}{M^2_a} = \pi \, K_a \frac{H}{M_a} \approx
 \frac{ K_a^3 M_a}{10^{19}\, \mathrm{GeV}}  \frac{T^2}{T^2_a}
	\; .
\eeq
These have to be small at the relaxation temperatures $T_a$ in order that the neutrino densities evolve as described in this section.
For decay constants $K_a$ of the order of 50 (100) this happens for neutrino masses $M_a$ below 
$10^{12}$ ($10^{11}$) GeV.
Then, the thermal masses can be neglected provided that the vacuum masses are not degenerate.

There is another point. 
The neutrino states produced in collisions are linear combinations of the mass eigenstates $N_a$ of the form $h_{ia} N_a$.
In a free path the mass eigenstate wave functions oscillate with unequal frequencies, 
$\sqrt{\mathbf{p}^2 + M_a^2}$, for the same linear momentum $\mathbf{p}$.
They differ by $(M_a^2 - M_b^2)/2 E$ 
for relativistic particles (energy $E\approx p$) and in a mean free path $1/\Gamma_a$
 give rise to average phase differences equal to
\be{phase}
\frac{M_a^2 - M_b^2}{2 \, \Gamma_a} \left\langle E^{-1} \right\rangle 
%%\left\langle \frac{1}{E} \right\rangle  
\approx 5 \frac{M_a^2 - M_b^2}{m^2_{aa}}
	\; ,
\eeq
where we made use of \eqs{Ta} and (\ref{mm}) with $\beta \approx 1/7$.
This shows that as long as thermal masses are much smaller than the vacuum mass gaps the neutrino mass eigenstates develop large phase decoherence between collisions so that one can consider that the neutrino states with definite number densities are the vacuum mass eigenstates $N_a$.
This is important because if, for example, the neutrino densities $n_a$ were all equal to each other the leptogenesis sources would vanish in the ultrarelativistic regime at lowest order, as the results of next sections show. 

%%\newpage
%%%%%%%%%																															%%%%%%%%%%%%%%%%%%%
%%%%%%%%%																															%%%%%%%%%%%%%%%%%%%
%%%%%%%%%																															%%%%%%%%%%%%%%%%%%%
%%%%%%%%%																															%%%%%%%%%%%%%%%%%%%
\subsection{Standard lepton asymmetries} \label{lepton}

%fffffffffffffffffffffffffffffffffffffffffffffffffffff
%fffffffffffffffffffffffffffffffffffffffffffffffffffff
\begin{figure}[b]
\begin{center}
%%\epsfbox{NHLFig1.eps}
\includegraphics[width=70mm]{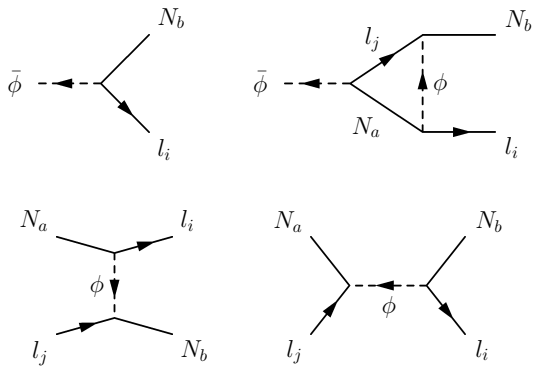}% Here is how to import EPS art
\end{center}
\caption{\label{fig1} Diagrams contributing to CP-asymmetries in decays, inverse decays and scatterings.}
%%\end{center}
\end{figure}
%fffffffffffffffffffffffffffffffffffffffffffffffffffff
%fffffffffffffffffffffffffffffffffffffffffffffffffffff

Leptogenesis is dominated by the following $CP$ asymmetric reactions~\cite{beyond03}:
Higgs decays into leptons and singlet neutrinos, inverse decays and scatterings of leptons off neutrinos.
The $CP$ asymmetries result from the diagrams of \fig{fig1}, more specifically, from the interference between the tree level amplitude and the absorptive part of the one-loop amplitude of the Higgs boson decay (inverse decay) and from the interference between the $t$ channel amplitude and the absorptive part of the $s$ channel amplitude of neutrino-lepton scattering.
The particle asymmetry sources vanish in thermal equilibrium 
but, as long as neutrinos stay rarefied scatterings and inverse decays do not match Higgs decays and particle asymmetries develop in the various lepton flavors, singlet neutrinos, and Higgs boson as well as enforced by hypercharge conservation.

The source terms (labeled with S) responsible for the generation of the standard family lepton numbers $L_i$ are given in leading order by
\bea{li}
(\dot{L}_i )_\s = \sum_{b} \, 
\dgamma (\barphi \ra N_b l_i)_\s - \dgamma (N_b l_i \ra \barphi)_\s
 \nonumber \\
 + \sum_{abj}  
\dgamma (N_a l_j\ra  N_b l_i )_\s - \dgamma (N_b l_i \ra N_a l_j)_\s 
	\; ,	
\eea
%%(\ref{li}) %%%%%%%%%%%%%%%%%%%%%%%%%%%%%%%%%%%%%%%%%%%%%%%%%%%%%%%%%%%%%%
where
\be{deltagamma} 
\dgamma (X \ra Y) = \gamma (X \ra Y) -\gamma (\bar{X} \ra \bar{Y})
\eeq
denotes the difference between the rate of an arbitrary reaction $X \rightarrow Y$ integrated over a fixed comoving volume 
and the rate of the conjugate reaction $\bar{X} \ra \bar{Y}$,
where $\bar{X}$ and $\bar{Y}$
are $CPT$ conjugate states of $X$ and $Y$ respectively 
($CPT$ transforms particles into antiparticles, reverses helicity but not momentum).
The difference $\dgamma (X \ra Y)$ can be separated in two types of contributions, the transport terms proportional to particle antiparticle abundance asymmetries and the source terms responsible for primordial asymmetry generation. 
By definition the source terms exist in absence of particle aymmetries and in the 
$\dgamma (X \ra Y)_\s$
expressions we assume that $CPT$ conjugate states have exactly the same distribution functions.

$CPT$ invariance and unitarity ensure that the total decay rates of any quantum state and its $CPT$ conjugate state are equal to each other.
This translates in constraints on the rate asymmetries:
\be{CPT}
\sum_Y \dgamma(X \ra Y)_S = 0
	\; .
\eeq
In leading order, one gets:
\be{CPTa}
\dgamma (N_b l_i \ra \barphi)_\s + 
\sum_{aj} \,  \dgamma (N_b l_i \ra N_a l_j)_\s =0
 	\;, 
\eeq
\be{CPTb}
\sum_{bi} \, \dgamma (\barphi \ra N_b l_i)_\s =0
 	\;.		
\eeq
%%(\ref{CPT}) %%%%%%%%%%%%%%%%%%%%%%%%%%%%%%%%%%%%%%%%%%%%%%%%%%%%%%%%%%%%%
Then, the $L_i$ source terms of \eq{li} can be written as,
\bea{lis}
(\dot{L}_i )_\s =  
\sum_{b}  \dgamma (\barphi \! \ra \! N_b l_i)_\s 	%%\nonumber	\\ 
+  \sum_{abj}  \dgamma (N_a l_j  \! \ra  \! N_b l_i )_\s .%%	\;.	
\eea
%%(\ref{lis}) %%%%%%%%%%%%%%%%%%%%%%%%%%%%%%%%%%%%%%%%%%%%%%%%%%%%%%%%%%%%%%

On the other hand, $CPT$ invariance implies that any reaction $X \ra Y$ has the same transition probability as $\bar{Y} \ra \bar{X}$ where the bars indicate $CPT$ conjugate states.
As a result the rate asymmetries 
$\dgamma (X \ra Y)$
obey the following constraints in thermal equilibrium assuming identical particle and antiparticle distribution functions (zero chemical potentials):
\be{CPTgeneral}
\dgamma (X \ra Y)_\equi + \dgamma (Y \ra X)_\equi = 0
	\; .
\eeq
Here one employs the identity~\cite{Weinberg79,Kolb80}
 satisfied by the thermal distribution functions of particles involved in two inverse reactions 
$X \ra Y$ and $Y \ra X$, 
schematically, 
$f_X (1 \pm f_Y) =  f_Y (1 \pm f_X)$,
where $f_X$ ($f_Y$) stands for the product of initial state particle distribution functions and 
$1 \pm f_Y$ ($1 \pm f_X$) for the product of final state stimulated emission and/or Pauli blocking factors. 
Applying the above constraint on \eqs{CPTa} and (\ref{CPTb})
one obtains 
\bea{CPTequi}
\dgamma (\barphi \ra N_b l_i)_\equi  + 
\sum_{aj} \dgamma (N_a l_j\ra  N_b l_i)_\equi = 0  , %%	\;,
		\label{CPTc} \\
\sum_{bi} \dgamma (N_b l_i \ra \barphi)_\equi =
\sum_{abij}  \dgamma (N_a l_j \ra N_b l_i)_\equi = 0 . %%	\;
				\label{CPTh}
\eea
This ensures that \eq{lis} satisfies the well known property that the asymmetry source terms vanish in thermal equilibrium~\cite{Weinberg79,Kolb80}.

The diagrams of \fig{fig1} lead to the following results:
\bea{liasy}
 \dgamma (\barphi \ra N_b l_i)_\s =	&		\nonumber \\ %%twocolumn
 \sum_{aj} J_{ijab} \, \int d\Phi  & 
 f_\phi  F_i F_j F_a F_b  
\frac{-4 p'_i \cdot p'_j}{(p_a-p_i)^2 - m^2_\phi}
 \; , \label{lidasy}		\\
  \dgamma (N_a l_j\ra  N_b l_i )_\s =	&	\nonumber \\ %%twocolumn
  J_{ijab} \, \, \int d\Phi  & 
  f_a f_j F_\phi F_i F_b 	 
\frac{4 p'_i \cdot p'_j}{(p_a-p_i)^2 - m^2_\phi}
 \; , \label{lisasy}
\eea
where $f_\alpha$ are particle distribution functions, $F_\alpha$ are fermion Pauli blocking factors $1- f_\alpha$ or the stimulated emission factor $F_\phi = 1+f_\phi$ and 
\be{dphi}
d \Phi = V 
\prod_{\alpha} \frac{d^3 p_\alpha}{(2\pi)^3 2 E_\alpha}  
	(2\pi)^8 \delta(p_a + p_j - p_\phi) \,  \delta(p_b + p_i - p_\phi)
 \;  
\eeq
is the phase space element running over all particles 
$\phi$, $l_i$, $l_j$, $N_a$, $N_b$ ($V$ is the spatial volume).
Standard leptons and Higgs boson are assumed to be in thermal equilibrium while singlet neutrinos have densities given by \eq{na} and distribution functions  proportional to the thermal distribution functions as in \eq{fa}.
The chemical potentials are equal to zero.
The factors 
\be{J}
J_{ijab} = \mathrm{Im} \{ h_{ia} h_{ja} h_{ib}^\ast h_{jb}^\ast  \} M_a M_b
\eeq
signal the necessary $CP$ and lepton number violation through Yukawa couplings and Majorana masses.
$J_{ijab}$ are antisymmetric under neutrino flavor exchange $a \leftrightarrow b$ and symmetric under lepton flavor exchange $i \leftrightarrow j$.
	The internal product 
$p'_i \cdot p'_j = p_i p_j - \mathbf{p}_i \cdot  \mathbf{p}_j$ 
in the integrand functions refers to pseudo 4-vectors $p'_i$.
The $p'_i$ space components coincide with the 3-vector linear momentum $\mathbf{p}_i$ 
and its time component is equal to the momentum absolute value, 
$p_i^{\prime 0} = p_i$.
	The 4-vectors $p'_i$ result from the spinor wave functions in a thermal bath~\cite{Weldon82a}: 
$\sum u\, \bar{u} =  \slax{p}'_i$.
	In contrast, the time component of the 4-momentum $p_i$ is the lepton energy, related with the lepton thermal mass as usual:
$p_i^0 = (p_i^2 + m_i^2)^{1/2}$.

The generation rate of total standard lepton number $L = \sum L_i$ is obtained from \eq{lis}.
The constraints of \eqs{CPTa} and (\ref{CPTb}) indicate that only scatterings, or better, inverse decays, contribute as  $L$ sources: 
\bea{Lls}
(\dot{L} )_\s =
\sum_{abij} \, \dgamma (N_a l_j\ra  N_b l_i )_\s = 
& 	\nonumber \\ %%twocolumn
 - \sum_{aj} \dgamma (N_a l_j \ra \barphi)_\s 
	\;.	
	&  %%twocolumn
\eea
%%(\ref{Lls}) %%%%%%%%%%%%%%%%%%%%%%%%%%%%%%%%%%%%%%%%%%%%%%%%%%%%%%%%%%%%%%
We evaluate the asymmetries including Pauli blocking effects and nonzero lepton thermal masses.
However, we neglect the neutrino masses inside the integrals because we are calculating the leading order contributions at relativistic temperatures $T \gg M_a$.
Neutrino masses appear in the constant factors $J_{ijab}$.
All the dependence on the neutrino flavors goes in $J_{ijab}$ and in the neutrino number densities $n_a$ that are functions of the temperature and flavor dependent relaxation temperatures $T_a$.
Notice that the following identity holds for the distribution functions in \eq{lisasy}
\be{fff}
f_a f_j (1 + f_\phi) = \frac{n_a}{n_\equi} f_\phi (1 - f_a^\equi) (1- f_j)
	\; .
\end{equation}
Under these conditions the generation rate of standard total lepton number is
\bea{Llsresult}
(\dot{L} )_\s
=  - \frac{c}{(8 \pi)^4} T^2 \, V
\sum_{abij} \frac{n_a - \nequi}{\nequi} J_{ijab}
	\;,
\eea
with $c \sim 3 $ ($c = 2.6$ for a Higgs thermal mass $m_\phi = 0.6 \,T$ and lepton thermal square masses $m^2_l = 0.036 \, T^2$).
It is clear that leptogenesis ceases when neutrinos reach thermal equilibrium abundances.

%%%%%%%%%																															%%%%%%%%%%%%%%%%%%%
%%%%%%%%%																															%%%%%%%%%%%%%%%%%%%
%%%%%%%%%																															%%%%%%%%%%%%%%%%%%%
%%%%%%%%%																															%%%%%%%%%%%%%%%%%%%
\subsection{Neutrino asymmetries} \label{neutrino}

In the above equations one sums over both $N_a$ neutrino helicities.
But as emphasized in this paper the leptogenesis processes generate also asymmetries in the abundances of the two helicity states $N_a^+$ and $N_a^-$.
The helicity asymmetries are denoted as $\Lambda_a$ and are defined with respect to the
cosmological comoving frame.
In the following the rate asymmetries like 
$\Delta \gamma (X \ra N_a^+ Y)$
denote the difference between the rates of the $CPT$ conjugate reactions
$X \ra N_a^+ Y$ and $\bar{X} \ra N_a^- \bar{Y}$.
Wherever the neutrino helicity does not appear explicitly a sum over helicities is assumed.
The leading order source terms are
\bea{lb}
 (\dot{\Lambda}_b)_\s =
\sum_{i}  
  \left\{
\dgamma (\barphi \ra N_b^+ l_i) - \dgamma (N_b^+ l_i \ra \barphi) 
  \right.
 \nonumber \\ 
 +  \left.
\dgamma (\phi \ra N_b^+ \bar{l}_i) - \dgamma (N_b^+ \bar{l}_i \ra \phi) %% +
  \right\}_\s
   \nonumber \\ 
 + \sum_{aij}  
  \left\{
\dgamma (N_a l_j\ra  N_b^+ l_i ) - \dgamma (N_b^+ l_i \ra N_a l_j)
  \right.
 \\ \nonumber
 +  \left.
\dgamma (N_a \bar{l}_j \ra  N_b^+ \bar{l}_i ) - 
\dgamma (N_b^+ \bar{l}_i \ra N_a \bar{l}_j)
  \right\}_\s
	\;.	
\eea
%%(\ref{lb}) %%%%%%%%%%%%%%%%%%%%%%%%%%%%%%%%%%%%%%%%%%%%%%%%%%%%%%%%%%%%%%
When $a=b$ in the second summation one gets the correct factor of $2$ for the 
 helicity flip reactions $N_b^\mp \ra N_b^\pm$.

The $\Lambda_b$ source terms are subject to $CPT$ invariance and unitarity conditions namely
\bsub
\dgamma (N_b^+ l_i \ra \barphi)_\s +  
\sum_{aj}  \dgamma (N_b^+ l_i \ra N_a l_j)_\s = 0		, %%	\;,
		\label{CPTd} \\  
\dgamma (N_b^+ \bar{l}_i \ra \phi)_\s +
\sum_{aj}  \dgamma (N_b^+ \bar{l}_i \ra N_a \bar{l}_j)_\s =0		.%%	\;.
 	\label{CPTe}
\esub
These constraints eliminate four of the source terms contained in \eq{lb}:
\bea{lbs}
 (\dot{\Lambda}_b)_\s =
\sum_{i}  
\dgamma (\barphi \ra N_b^+ l_i)_\s 
+ \dgamma (\phi \ra N_b^+ \bar{l}_i)_\s
 \nonumber \\
 + \sum_{aij}  
\dgamma (N_a l_j\ra  N_b^+ l_i )_\s +
\dgamma (N_a \bar{l}_j \ra  N_b^+ \bar{l}_i)_\s		.%%	\;.	
\eea
%%(\ref{lbs}) %%%%%%%%%%%%%%%%%%%%%%%%%%%%%%%%%%%%%%%%%%%%%%%%%%%%%%%%%%%%%% 
From the $CPT$ invariance condition under thermal equilibrium and zero chemical potentials, \eq{CPTgeneral}, one derives also that
\bsub
\dgamma (\barphi \ra N_b^+ l_i)_\equi +  
\sum_{aj}  
 \dgamma (N_a l_j \ra N_b^+ l_i)_\equi = 0 	,%%	\;,
		\label{CPTf} \\  
\dgamma (\phi \ra N_b^+ \bar{l}_i)_\equi +
\sum_{aj}
 \dgamma (N_a \bar{l}_j \ra  N_b^+ \bar{l}_i)_\equi = 0		,%%	\; ,
	\label{CPTg}
\esub
which ensures that the $\Lambda_b$ source terms vanish in thermal equilibrium.

The rate asymmetries in \eq{lis}  contain by definition sums over neutrino helicity states and therefore relate to the ones of \eqs{lbs}  as
\bsub
\dgamma (\barphi \ra N_b l_i)  & = &
\dgamma (\barphi \ra N_b^+ l_i)
 		\nonumber \\   	%%twocolumn
& &  - \dgamma (\phi \ra N_b^+ \bar{l}_i)		
	\;,	  	%%twocolumn
	\label{asydecay} \\  
\dgamma  (N_a l_j \ra  N_b l_i ) & = &  
\dgamma (N_a l_j\ra  N_b^+ l_i ) 
 		\nonumber \\   	%%twocolumn
& & - \dgamma (N_a \bar{l}_j \ra N_b^+ \bar{l}_i)
	\;. 	  	%%twocolumn	
	\label{asyscatt}
\esub
Replacing this in \eqs{lis} and (\ref{lbs}) one obtains the leading source terms of the total 'lepton number' $L'=\sum L_i + \sum \Lambda_a$:
\bea{lprimes}
 (\dot{L}')_\s  &= &
2 \sum_{bi}  
\dgamma (\barphi \ra N_b^+ l_i)_\s   \nonumber \\
& + & 2 \sum_{abij}  
\dgamma (N_a l_j \ra  N_b^+ l_i)_\s  
	\;.	
\eea
%%(\ref{lprimes}) %%%%%%%%%%%%%%%%%%%%%%%%%%%%%%%%%%%%%%%%%%%%%%%%%%%%%%%%%%%%%%

	The asymmetries are calculated from the diagrams of \fig{fig1}.
	The $N_b$ positive (negative) helicity states are specified with spinors satisfying \eq{ub}.
	The results are the following, using the same notations as in 
\eqs{lidasy}-(\ref{dphi}):
\bea{lsource}
 \dgamma (\barphi \ra N_b^+ l_i)_\s = 	&		\nonumber \\ %%twocolumn
 \sum_{aj} J_{ijab}  \int d\Phi 				&	  %%twocolumn
  f_\phi  F_i F_j F_a F_b   \frac{ - 2 \tau }{(p_a-p_i)^2 - m^2_\phi}
 \; , \label{lasydecay}		\\
  \dgamma (N_a l_j\ra  N_b^+ l_i )_\s = 	&		\nonumber \\ %%twocolumn
  J_{ijab}  \int d\Phi 									&	  %%twocolumn
   f_a f_j F_\phi F_i F_b 	\frac{ 2 \tau}{(p_a-p_i)^2 - m^2_\phi}
 \; , \label{lasyscatt}
\eea
with
\be{tau}
\tau = p'_i \cdot p'_j + \frac{p_i}{p_b}\, p_b \cdot p'_j - \frac{p_j}{p_b}\, p_b \cdot p'_i
	\; .
\eeq
Notice that the $N_b$ helicity states as well as the absolute 3-momenta $p_i$, $p_j$, $p_b$ in the expressions above are defined with respect to the cosmological comoving frame.
As before we neglect the neutrino masses inside the phase space integrals.
The integrand function $\tau$ is finite however, if one neglects lepton thermal masses and Pauli blocking factors the integration over the fermion angular variables yields a null result
(recall that the distribution function factor in \eq{lasyscatt} obeys the relation (\ref{fff})).
This unexpected result means that in that approximation $L'$ is not generated at all i.e., the total asymmetry, $\Lambda = \sum \Lambda_b$, generated in the neutrino sector
cancels exactly the total lepton number generated in the standard lepton sector.
But for nonzero lepton thermal masses the result is finite:
\bea{L'sresult}
(\dot{L}')_\s
=  \frac{c}{(8 \pi)^4} T^2 \, V
\sum_{abij} \frac{n_a - \nequi}{\nequi} J_{ijab }
	\;,	
\eea
with $c \sim  1 $ ($c = 1.2$ for Higgs thermal mass $m_\phi = 0.6 \,T$ and lepton thermal square masses $m^2_l = 0.036 \, T^2$).

The comparison with \eq{Llsresult} shows that leptogenesis generates a total neutrino helicity asymmetry $\Lambda$ of opposite sign and larger than the standard lepton asymmetry $L$ so that the source of $L' = L + \Lambda$ is of opposite sign to $L$.
The neutrino and lepton asymmetries are not separately conserved.
They are both violated by neutrino Yukawa couplings and the lepton number $L$ is violated by weak sphalerons as well.
These interactions cause the interchange of $B-L'$ between the singlet neutrino and standard lepton and quark sectors but conserve $B-L'$ in absence of neutrino Majorana masses.
The neutrino masses are so responsible for the $B-L'$ dissipation. 
In next section we study this effect and establish the evolution of $B-L'$.

%%\newpage
%%%%%%%%%																															%%%%%%%%%%%%%%%%%%%
%%%%%%%%%																															%%%%%%%%%%%%%%%%%%%
%%%%%%%%%																															%%%%%%%%%%%%%%%%%%%
%%%%%%%%%																															%%%%%%%%%%%%%%%%%%%
\section{Washout processes} \label{washout}

Weak sphalerons and neutrino Yukawa couplings are the only ones that violate the standard partial lepton numbers $L_i$.
The processes like 
$\barphi \ra N_a l_i$, $\bar{t}q_t \ra N_a l_i$
violate the standard total lepton number $L = \sum L_i$ by one unit, but this does not mean that they dissipate the total lepton number that is generated in leptogenesis.
Indeed, one has to distinguish between the two neutrino helicity states $N_a^{\pm}$.
The reactions
$\barphi \ra N_a^- l_i$, $\,\bar{t}\, q_t \ra N_a^- l_i$, $\, \barphi \,A \ra N_a^- l_i$
($A$ is a gauge boson)
$N_a^+ \barphi \ra l_i A$, $\, N_a^+ A\ra l_i  \phi$, 
$\,N_a^+ \bar{t} \ra \bar{q_t}\, l_i$, $\,N_a^+ q_t \ra t\, l_i$
violate $L$ by one unit but conserve the total 'lepton' number $L' = L + \Lambda$
because they decrease the total helicity $\Lambda = \sum \Lambda_a$ by one unit.
On the contrary, the same processes with opposite helicity neutrino states violate $L'$ by two units.
They require nonzero neutrino Majorana masses and are suppressed at relativistic temperatures 
$T \gg M_a$. 
The $L'$ conserving processes dominate the neutrino thermalization and enter in equilibrium with relaxation temperatures $T_a$ given by \eq{Ta}.
Below these temperatures the $L'$ conserving processes are fast in comparison with the expansion rate and thus enforce constraints on the lepton and neutrino chemical potentials:
\be{etaphilN}
\eta_\phi + \eta_{i} - \eta_a = 0
	\;.
\eeq
This implies that due to rapid flavor violation all lepton doublets $l_i$ have equal degeneracy parameters $\eta_i = \eta_l$ and all positive (negative) helicity neutrinos 
$N_a^+$ ($N_a^-$) have equal degeneracy parameters 
$\eta_a = \eta_\phi + \eta_l$ ($-\eta_a$).
As a result the partial asymmetries $L_i$ and $\Lambda_a$ are not directly related with the respective leptogenesis sources but are rather determined from the overall $B-L'$ asymmetry by the set of chemical potential constraints.

There are as much constraints as independent fast flavor changing reactions and this depends on the temperature scale~\cite{Khlebnikov88,Harvey90}.
In the standard model case the latest interactions to enter in equilibrium are the Yukawa couplings and their equilibrium temperatures depend on the particular right-handed quark or lepton isosinglet
(for a detailed discussion see refs.~\cite{beyond03,Bento03}).
In any case the constraints leave $B-L'$ as a free variable and determine the other quantum numbers, in particular $L_i$ and $\Lambda_a$, as proportional to $B-L'$.
In turn, $B-L'$ is completely determined by the leptogenesis sources on one hand and the dissipation (washout) processes on the other hand.

$B-L'$ is violated by two units in the $\Delta L = 1$, $\Delta \Lambda =1$ reactions 
$\barphi \ra N_a^+ l_i$, $\,\bar{t}\, q_t \ra N_a^+ l_i$, $\, \barphi \,A \ra N_a^+ l_i$,
and crossed channels
%% $\,N_a^- \barphi \ra l_i A$, $\, N_a^- A\ra l_i  \phi$, 
%%$\,N_a^- \bar{t} \ra \bar{q_t}\, l_i$, 
$\,N_a^- q_t \ra t\, l_i$, ...,
in the $\Delta L = 2$, $\Delta \Lambda =0$ scatterings
$\barphi \, \barphi \ra l_i  l_j$, $\,\barphi \, \bar{l}_j \ra \phi \, l_i $, 
$\,\bar{l}_i \bar{l}_j \ra \phi \, \phi$, 
and in the $\Delta L = 0$, $\Delta \Lambda =2$ scatterings
${l}_i \bar{l}_j, \phi \, \barphi \ra N_a^+ N_b^+$, 
$\, \phi \, N_b^- \ra \phi \, N_a^+$, 
$\, {l}_j \, N_b^- \ra {l}_i \, N_a^+$,
$\, \bar{l}_j \, N_b^- \ra \bar{l}_i \, N_a^+$,
$N_a^- N_b^- \ra {l}_i \bar{l}_j, \phi \, \barphi$.
All these processes depend on the existence of neutrino Majorana masses.
In the $\Delta L = 2$ case they contribute through the neutrino propagators and in the 
$\Delta L = 0,1$ reactions they make possible that a neutrino be produced or annihilated with the 'wrong' helicity state i.e., helicity opposite to the chirality determined by the Yukawa couplings $l_i N_a \phi$. 
The left-handed (right-handed) chiral projection of a positive (negative) helicity state with energy $E$ goes as $M_a/2E$ in the relativistic limit.
Thus, the $L'$ violating processes are suppressed with respect to the respective $L'$ conserving channels by a ratio going as $M_a^2/4E^2$.
The thermal average of this ratio is about $M_a^2/16 T^2$ for scatterings but in the case of 
$\barphi \ra N_a l_i$ decays the center of mass energy, $m_{\phi} \sim 0.6 \,T$, and average neutrino energy are much smaller.
The $\barphi \ra N_a^+ l_i$ branching fraction is given by
\be{Ba}
B_{a}^+ = \frac{\gamma(\barphi \ra N_a^+ l_i)}{\gamma(\barphi \ra N_a l_i)}
\approx \frac{8 M_a^2}{T^2} 
	\; ,
\eeq
two orders of magnitude higher than in the case of scatterings.
	As the $\Delta L = 1$ scatterings 
%%$\bar{t}\, q_t \ra N_a l_i$, $\, \barphi \,A \ra N_a l_i$, ...
have total rates comparable with the Higgs decays, this means that 
the Higgs decays dominate the $L'$ violating rates by two orders of magnitude over the scatterings.
	As far as $\Delta L = 0$ and $\Delta L = 2$ scatterings is concerned,
$\phi \, \barphi \ra N_a^+ N_b^+$, ..., $\barphi \, \barphi \ra l_i  l_j$, ..., 
they are also suppressed by the same $10^{-2}$ factor plus an extra factor of $|h_{ia}|^2$ with respect to the $\Delta L' =2$ Higgs decays,
because they get one more Yukawa coupling than the $\Delta L = 1$ scatterings.

	From the discussion above one learns that the Higgs decays and inverse decays dominate the violation of $B-L'$.
Hence, the evolution of $B-L'$ is well described by the equation
\bea{bmla}
\dot{B}-\dot{L}'   =  - (\dot{L'})_\s  +	 
 2   \sum_{ai} \{ 
\dgamma (N_a^+ l_i \ra \barphi)_0 			\nonumber \\ %%twocolumn
- \dgamma (\barphi \ra N_a^+ l_i)_0 
 \} 	\;,	 
\eea
where the label '0' indicates that the rates are calculated at tree level contrary to the leptogenesis source term $(\dot{L'})_\s$ given in \eq{lprimes}.
At temperatures below $T_a$ neutrinos have thermal equilibrium abundances and the transport terms depend on the chemical degeneracy parameters as follows:
\be{bmlb}
\dot{B}-\dot{L}' =  - (\dot{L'})_\s + 4 \sum_{ai} \, 
\gamma_{\barphi \ra N_a^+ l_i} (\eta_\phi + \eta_i + \eta_a) 
	\; .	
\eeq
Moreover, the $L'$ conserving neutrino reactions are in equilibrium under $T_a$ and the constraint of \eq{etaphilN} applies, so that 
$\eta_i = \eta_l$ and $\eta_a = \eta_\phi +\eta_l$.
The degeneracy parameters parametrize the particle antiparticle number asymmetries.
For example, for a ultrarelativistic fermion $f$ the asymmetry is given in leading order by
$Y_f - Y_{\bar{f}}= 1.83 \, \eta_f Y_\equi$~\cite{Kolb80,Kolb90}.
The constraints imposed by the standard model interactions leave one degeneracy parameter as free parameter~\cite{Khlebnikov88,Harvey90}. 
In our case they lead to a relation of the form
\be{r} 
B-L' = - r (\eta_\phi + \eta_l) Y_\equi
	\; .
\eeq
The precise value of the factor $r$ depends on the set of chemical constraints and therefore on the temperature scale.
It varies from $r = 12.8$ at temperatures above $10^{12} \gev$, where the Yukawa couplings of the isosinglet right-handed $b_R$ quark and $\tau_R$ lepton are not yet in equilibrium, to 
$r = 18.6$ under $10^{4} \gev$, where all quark and lepton Yukawa couplings are in equilibrium.

The $\barphi \ra N_a^+ l_i$ partial decay rate is a fraction of the $\barphi \ra N_a l_i$ decay rate
 as indicated in \eq{Ba}.
On the other hand, the Higgs decay rates scale with the first power of the temperature and can be related in the same way as in \eq{Ta} with the Hubble expansion rate $H$:
\be{Taprime}
\sum_i \gamma_{\barphi \ra N_a l_i} = \frac{T'_a}{T} H \, Y_\equi
	\; .
	\eeq
The temperatures $T'_a$ are about one half smaller than the relaxation temperatures $T_a$ because unlike the later $T'_a$ do not receive contributions from scattering processes.
Combining everything one rewrites \eq{bmlb} in the form,
\be{bmlc}
\frac{d(B-L')}{d T^{-3}} =  
- \left( \frac{d L'}{d T^{-3}}\right)_\s  - \tstar^3 (B-L')
	\; ,	
\eeq
\be{tin}
\tstar^3 = \frac{8}{3r} \sum_a B_a^+ T^2 T'_a \sim \frac{2}{3} \sum_a M_a^2 T_a  
	\; .
\eeq
This shows that when leptogenesis processes are ineffective, at temperatures $T < T_a$, $B-L'$ decays exponentially with $T^{-3}$ and damping constant equal to $\tstar^3$. 
The same happens with the neutrino degeneracy parameters 
$\eta_a = \eta_\phi + \eta_l$,
 as determined by \eq{r}.

It is important to recognize the difference between $\tstar$ and the relaxation temperatures $T_a$.
	At $T_a$ neutrino reactions enter in thermal equilibrium  and the lepton and neutrino partial quantum numbers $L_i - B/3$ and $\Lambda_a$ begin to be rapidly violated.
	However, it does not mean that the standard lepton number $L$ or $B-L'$ start to be washed out at $T_a$.
	It only means that the effective couplings $l_i N_a^- \phi$ enter in equilibrium and the constraints $\eta_\phi + \eta_{i} - \eta_a = 0$
are enforced.
The complete set of chemical equilibrium constraints force $B-L'$ to distribute into $B$, $-L$ and $-\Lambda$ in similar proportions.
On the other hand the effective couplings $l_i N_a^+ \phi$ are not yet in equilibrium because the temperatures $T_a$ are much larger than the neutrino Majorana masses.
These couplings enter in equilibrium later at the temperature $\tstar$.
Below that temperature the constraints 
$\eta_\phi + \eta_{i} + \eta_a = 0$
should apply on top of the previous constraints 
$\eta_\phi + \eta_{i} - \eta_a = 0$.
It means that the degeneracy parameters $\eta_a$, $\eta_\phi + \eta_{i}$, and $B-L'$ 
as well as all asymmetries proportional to $B-L'$ are strongly damped below the temperature $\tstar$.
But not above $\tstar$.
In fact, $B-L'$ is only marginally damped at the $T_a$ temperature scale.

For decay constants $K_a$ as large as 70, neutrinos enter in equilibrium at temperatures 
$T_a  \approx \frac{1}{7} K_a M_a \sim 10\, M_a$
but the $B-L'$ damping constant $\tstar^3$ is two orders of magnitude smaller than the $T_a^3$ scale: 
$\tstar^3 \sim \frac{1}{150} \sum_a T_a^3$.
Moreover, the greater the decay constants $K_a$ are the smaller is $\tstar^3$ in comparison with $T_a^3$. 
On the other hand $\tstar$ is quite close to the neutrino mass scale.

This gives an important lesson.
If one ignores neutrino helicity asymmetries, as has ever been done in the literature, then one obtains that the temperatures $T_a$ set the chemical potential constraints
$\eta_i + \eta_\phi =0$
and define the $B-L$ relaxation temperature scale.
	This is wrong.
	The correct constraints must include the neutrino degeneracy parameters $\eta_a$ and are given by \eq{etaphilN}.
Ignoring neutrino helicity asymmetries leads to an overestimate by two orders of magnitude of the damping rate of the lepton and quark asymmetries.

So far we have considered that singlet neutrinos have equilibrium temperatures 
$T_a \gtrsim 10\, M_a $ 
larger than any of their masses $M_a$.
	Another scenario that has been often considered for simplicity is the hierarchical 
scenario
\cite{Buchmuller00,Fukugita86,Luty92,Flanz98,Hirsch01,Buchmuller02,Giudice04}
where one of the singlet neutrinos, $N_1$, is much lighter than the others.
	Then, it is argued that any asymmetries generated at the heaviest neutrinos decaying phases are later washed out by the lightest neutrino $\Delta L = 1$ reactions before the temperature reaches the mass $M_1$.
	As we have just shown this is not correct because it does not take into account the $N_1$ helicity asymmetry. 
	One concludes from \eq{tin} that the quantum number $B-L'$ and all particle asymmetries related to it by chemical equilibrium constraints are damped with a relaxation temperature given by
$\tstar^3 \approx \frac{2}{3} M_1^2 T_1$
(in cases where the $\Delta L =2$ reactions mediated by off-shell neutrinos are not significant).
$\tstar$ is quite close to the mass $M_1$ which means that the asymmetries left after the heavy neutrino decays remain conserved at temperatures above the lightest neutrino mass.

Another possibility is that the lightest neutrino enters in equilibrium at a relaxation temperature $T_1$ below the other neutrinos masses.
 Then, the helicity asymmetry $\Lambda_1$ remains conserved when the heaviest neutrinos decay even if the standard lepton asymmetries are completely washout by their $\Delta L =1$ collisions or mediated $\Delta L =2$ reactions.
 Later on, the neutrino asymmetry $\Lambda_1$ is converted into a standard lepton asymmetry at temperatures below $T_1$.
 Preliminary numerical calculations~\cite{Bento05} indicate that as much as $1 \%$ of the neutrino asymmetry may be converted into a final $B-L$ asymmetry even if there is no $CP$ asymmetry associated with the lightest neutrino decay.
 This 
%% occurs in the parameter range %%
%% $\tilde{m}_1 = (h^\dagger h)_{11} {v^2}/{M_1} \lesssim 0.01$ eV and %%
makes a radical contrast with the traditional scenario and will be further investigated.

%%\newpage
%%%%%%%%%																															%%%%%%%%%%%%%%%%%%%
%%%%%%%%%																															%%%%%%%%%%%%%%%%%%%
%%%%%%%%%																															%%%%%%%%%%%%%%%%%%%
%%%%%%%%%																															%%%%%%%%%%%%%%%%%%%
\section{Conclusions}\label{conclusions}

We studied an aspect of the leptogenesis mechanism that has ever been overlooked.
We showed that as a rule the two helicity states of Majorana neutrinos do not have exactly the same abundances, contrary to what has been tacitly assumed.
The helicity asymmetries defined as differences between the two helicity state abundances of each neutrino species can be parametrized with appropriate neutrino chemical potentials as any other particles asymmetries.
The leptogenesis processes generate neutrino helicity asymmetries of the same order of magnitude as the standard lepton asymmetries.
This is quite natural because in the ultrarelativistic limit neutrino Majorana masses are negligible and the neutrino - lepton Yukawa couplings conserve a total lepton number assigned as +1 for right-handed neutrinos and -1 for left-handed neutrinos.

The neutrino helicity asymmetries participate in the system of Boltzmann equations that govern the evolution of particle asymmetries and do not decouple because for any particular reaction the two neutrino helicity states have distinct reaction rates.
	A reaction where a incoming left-handed lepton doublet goes with a incoming (outgoing) neutrino with positive (negative) helicity is suppressed with respect to the reaction where the same neutrino species is in the opposite negative (positive) helicity state.
The former is possible only because the neutrino has a Majorana mass but its amplitude is suppressed by the Lorentz contraction factor.
	The two reactions rates coincide only when the neutrino is at rest.
	As a result the standard lepton and neutrino sectors interchange asymmetries with each other which affects significantly the transport and dissipation of lepton number.
	
It proves convenient to work with the quantum number $B-L'$ where $L'$ is the sum of the total standard lepton number, $L$, and the total neutrino helicity asymmetry, $\Lambda$ 
(equal to the difference between the total number of neutrinos with positive helicity and the total number of negative helicity neutrinos).
$B-L'$ is conserved by sphalerons, its violation by Yukawa couplings is suppressed by the neutrino mass over temperature ratios, and reduces to the standard model $B-L$ number when the singlet neutrinos vanish from the Universe.
The chemical potential constraints enforced by the fast reactions at any given moment set the asymmetries of particles in equilibrium as proportional to $B-L'$.

We made a detailed analysis of the period when neutrinos are ultrarelativistic which permits simplifying approximations.
	It shows that one has to distinguish between the temperature $T_a$ at which a neutrino species $N_a$ comes into equilibrium and the temperature scale where its reactions have a significant damping (washout) effect on $B-L'$ and all particle asymmetries proportional to $B-L'$
($B-L$ in particular).
A neutrino species comes into equilibrium when its $\Delta L =1$ reactions become fast in Hubble rate terms.
In the existing literature all $\Delta L \neq 0$ reactions contribute to wash out the $B-L$ asymmetry.
But this is not right because it does not differentiate between the neutrino helicity states.
Some neutrino helicity configurations conserve $L'$.
These channels dominate the rates and exist even in absence of neutrino Majorana masses.
The other channels violate $L'$ and contribute to wash out $B-L'$ but they are subdominant and are suppressed by the second power of neutrino Lorentz contraction factors.
	A consequence of this is that the contribution of $\Delta L =1$ reactions to the $B-L'$ damping rate is two orders of magnitude smaller than expected if the neutrino helicity asymmetries are ignored.
	
In section \ref{generation} we studied the generation processes in the period of neutrino thermal production.
	It turns out that the $B-L'$ asymmetry generated directly in the neutrino helicity sector is of opposite sign and larger than the asymmetry generated in the standard lepton sector.
	We did not attempt to study the decaying phase of singlet neutrino(s).
	The calculations are more involved because the temperature is of the order of the neutrino mass(es).
	But one can identify one important difference with the traditional hierarchical scenario, where one of the singlet neutrinos is much lighter than the others:
the asymmetries left when the heaviest neutrinos vanish from the Universe are not necessarily washed out by the fast $\Delta L =1$ reactions of the lightest neutrino before the temperature reaches its mass scale.
	Moreover, even if the heaviest neutrino $\Delta L =1$ collisions or $\Delta L =2$ mediated reactions washout the standard lepton and Higgs boson asymmetries, the helicity asymmetry carried by the lightest neutrino may survive if it enters in equilibrium at a relaxation temperature below the heaviest neutrino masses.
	The neutrino asymmetry is later converted into a standard global $B-L$ asymmetry at temperatures close to its mass.
	This has been supported by numerical calculations and will be reported elsewhere~\cite{Bento05}.

%%\bigskip
\section*{Acknowledgments}
%%\ack{
This work was partially supported by the FCT grants 
CERN/FNU/43666/2001, %% and  \newline  
POCTI/FNU/43666/2002.  
%%CERN/FNU/43666/2001,    POCTI/FNU/43666/2002.  
%%}

%%\end{document}

%%\newpage
%%\Bibliography{99}

\end{document}